\RequirePackage{fix-cm}
\documentclass[twocolumn]{svjour3}          
\smartqed  

\usepackage[square]{natbib}

\usepackage{graphicx}
\usepackage{siunitx}
\usepackage[draft]{listofsymbols}
\usepackage{amsmath,amssymb}
\usepackage{shellesc}
\usepackage{tikz}
\usetikzlibrary{external}
\usepackage{pgfplots} 
\usepgfplotslibrary{external}
\usepackage{pgfgantt}
\usepackage{pdflscape}
\usepackage[export]{adjustbox}
\usepackage[permil]{overpic}
\usepackage{subfig}
\pgfplotsset{compat=newest} 
\pgfplotsset{plot coordinates/math parser=false}
\usepackage{verbatim} 
\usepackage{ marvosym }
\usepackage[normalem]{ulem}

\usetikzlibrary{shapes.arrows}

\usepackage[colorlinks,linkcolor=blue,citecolor=blue,urlcolor=blue,bookmarks=false,hypertexnames=true]{hyperref}

\usepackage[switch,columnwise]{lineno}

\newcommand{\uproman}[1]{\uppercase\expandafter{\romannumeral#1}}

\begin{document}

\sloppy

\title{On the calibration of Astigmatism particle tracking velocimetry for suspensions of different volume fractions
}
\subtitle{Applying APTV to suspensions flows}

\author{Philipp Brockmann \and Jeanette Hussong
}
 
 \institute{
 {Philipp Brockmann} (\Letter)   \and Jeanette Hussong \at
             Institute for Fluid Mechanics and Aerodynamics\\
             TU Darmstadt 
             \\ Flughafenstr. 19, 64347, Darmstadt-Griesheim \\ 
              \email{brockmann@sla.tu-darmstadt.de} \\ \\
 }

\date{Received: date / Accepted: date}

\maketitle

\begin{abstract}
\color{black}{
In the present study we demonstrate for the first time how Astigmatism Particle Tracking Velocimetry (APTV) can be utilized to measure suspensions dynamics. Measurements were successfully performed in monodisperse, refractive index matched suspensions of up to a volume fraction of $\Phi=19.9\%$.
For this, a small percentage ($\Phi<0.01\%$) of the particles is labeled with fluorescent dye acting as tracers for the particle tracking procedure. Calibration results show, that a slight deviation of the refractive index of liquid and particles leads to a strong shape change of the calibration curve with respect to the unladen case. This effect becomes more severe along the channel height. 
To compensate the shape change of the calibration curves the interpolation technique developed by Brockmann et al. (Experiments in Fluids, 61(2), 67, \citeyear{brockmann2020utilizing}) is adapted.
Using this technique, the interpolation procedure is applied to suspensions with different volume fractions of $\Phi<0.01\%$, $\Phi=4.73\%$, $\Phi=9.04\%$, $\Phi=12.97\%$, $\Phi=16.58\%$ and $\Phi=19.9\%$. To determine the effect of volume fraction on the perfomance of the method, the depth reconstruction error $\sigma_z$ and the measurement volume depth $\Delta z$, obtained in different calibration measurements, are estimated. 
Here, a relative position reconstruction accuracy of $\sigma_z$/$\Delta z$=0.90\% and $\sigma_z$/$\Delta z$=2.53\% is achieved for labeled calibration particles in dilute ($\Phi<0.01\%$) and semi-dilute ($\Phi\approx19.9\%$) suspensions, respectively. 
The measurement technique is validated for a laminar flow in a straight rectangular channel with a cross-sectional area of 2.55$\times$30\,mm$^2$. Uncertainties of 1.39\% and 3.34\% for the in-plane and 9.04\% and 22.57\% for the out-of-plane velocity with respect to the maximum streamwise velocity are achieved, at solid volume fractions of $\Phi<0.01\%$ and $\Phi=19.9\%$, respectively.
}
\end{abstract}

\section{Introduction}
\label{intro}
%
Particle laden flows are widely present in nature and technical settings such as sediment transport in rivers or industrial waste slurry transportation. In such flows, the particle-particle interaction and the hydrodynamic interaction between particles and fluid can give rise to various phenomena such as shear induced migration, inertial migration, shear thinning and shear thickening (Leighton and Acrivos \citeyear{leighton1987shear}, Ho and Leal \citeyear{ho1974inertial}, Stickel and Powell \citeyear{stickel2005fluid}). 
Both, numerical simulations and experiments are performed to reveal the underlying physical mechanisms behind these phenomena (Morita et al. \citeyear{morita2017equilibrium}, Fornari et al. \citeyear{fornari2018suspensions}, Shichi et al. \citeyear{shichi2017inertial}). Recent progress in numerical algorithms and computer technology enables researchers to examine the behavior of individual particles in great detail in suspensions even at high solid volume fractions (Kazerooni et al. \citeyear{kazerooni2017inertial}). 
Nonetheless, only a few experimental studies address the individual particle dynamics in suspensions beyond the dilute regime by means of optical measurement techniques such as Particle Image Velocimetry (PIV) (Zade et al. \citeyear{zade2018experimental, zade2019buoyant}, Zhang et al. \citeyear{zhang2018experimental}) or a combination of PIV and 2D Particle Tracking Velocimetry (PTV) (Baker and Coletti \citeyear{baker2019experimental}).
Optical measurement techniques have been proven to be reliable tools for investigating fluid flow problems with high spatial and temporal resolutions. However, when it comes to suspension flows at high solid volume fractions, they can not be utilized as the flow becomes opaque, given the presence of a large number of particles. 


A solution for this issue is refractive index matching (RIM) such that the resulting suspension becomes transparent (Wiederseiner et al. \citeyear{wiederseiner2011refractive}). As the required monodisperse particles as well as the RIM liquids are expensive and in most cases hazardous, it is desirable to perform the experiments in micro-scale setups. Micro-scale experiments in turn usually only provide restricted optical access, such that most three dimensional optical measurement techniques, which require complex calibration procedures, can not be employed. 
In this regard, Astigmatism Particle Tracking Velocimetry (APTV) appears to be a cost-effective and easy-to-implement method   for measuring the threedimensional motion of both liquid and particles.  

APTV is a single camera technique that can be applied to micro-scale enviroments with limited optical access (Cierpka et al. \citeyear{cierpka2012particle}). The basic idea of APTV is to introduce a controlled astigmatism to an optical system such that the particle's image deforms based on its depth position. Hence, the particle out-of-plane position can be reconstructed by the shape of the particle's image.
While the origin of APTV is found in micro-fluidic applications (Kao et al. \citeyear{kao1994tracking}), the technique was improved by various authors (Angarita Jaimes et al. \citeyear{angarita2006wavefront}, Chen et al. \citeyear{chen2009wavefront}, Rossi and Kähler \citeyear{rossi2014optimization}) and has been employed to investigate numerous physical phenomena at length scales ranging from \SI{9}{\micro\meter} (Huang et al. \citeyear{huang2016ultra}) up to \SI{5.1}{\milli\meter} (Buchmann et al. \citeyear{buchmann2014ultra}).
A detailed overview on APTV, its applications and its background is given by Cierpka and Kähler \citeyear{cierpka2012particle}.

Apart from utilizing APTV in combination with small tracer particles to measure the motion of liquids, in recent years the technique is more and more applied to measure the dynamics of the particles themselves in situations where particles undergo their own dynamics and do not follow the flow streamlines. 
Rossi et al. \citeyear{rossi2019particle} used APTV to measure particle velocity and concentration of an electrokinetically induced particle pattern in a suspension with a solid volume fraction of $\Phi=0.05\%$ inside a channel of 350$\times$\SI{30}{\square\micro\meter} crossection. They used a mixture of labeled particles with different dyes  ($d_p=\SI{0.245}{\micro\meter}$) with the majority of them being invisble during the APTV measurement to allow for higher solid volume fractions. They concluded that APTV is capable of measuring the concentration profile with a better resolution compared to classical segmentation approaches. 
Using APTV Blahout et al. \citeyear{blahout20203d} investigated the fractionation of $d_p=\SI{3.55}{\micro\meter}$ and $d_p=\SI{9.87}{\micro\meter}$ particles in a serpentine channel with a crossection of 200$\times$\SI{50}{\square\micro\meter}. They observed, that a transition of particles from four to two equilibrium trajectories takes place at particle size dependant bulk Reynolds numbers.

Most recent works increasingly focus on using machine learning tools such as deep neural networks to further increase the degree of automatization of APTV (Rossi and Barnkob \citeyear{rossi2019toward}) or to apply it on scenarios with low signal-to-noise ratios (Franchini et al. \citeyear{franchini2019calibration}).
König et al. \citeyear{koenig2020use} compared the performance of conventional and neural network supported APTV utilizing a bidisperse suspension in a laminar channel flow. While they achieved good results with both methods for the bidisperse suspension, they concluded that the neural network supported APTV is more robust against optical abberations and can be of great use for investigating suspensions with different particles sizes, shapes and even with particle clusters.

Efforts are also undertaken to further adapt and extend the classical APTV technique. 
Zhou et al. \citeyear{zhou2020holographic} introduced a modification utilizing Holographic imaging principles to establish Holographic Astigmatic Particle Tracking Velocimetry (HAPTV). Using a nozzle flow as a test case, they could successfully validate their measurement technique.
Brockmann et al. \citeyear{brockmann2020utilizing} developed an adapted APTV procedure, referred to as Ball Lens Astigmatism Particle Tracking Velocimetry (BLAPTV), for transparent particles that are large in comparison to the field of view. They extended the 2D Euclidean calibration approach developed by Cierpka et al. \citeyear{cierpka2010calibration} to the 3D space by additionaly considering the particles light intensity. Furthermore, they developed an interpolation method to account for shape changes of the calibration curve that occur in a channel along the channel height. The authors validated their proposed method by measuring the velocity profile of a laminar channel flow.

In the present study we combine conventional APTV with refractive index matching to investigate the dynamics of large particles ($d_p=\SI{60}{\micro\meter}$) for the first time by means of APTV in suspensions of up to $\Phi=19.9\%$ volume fraction. Here, a small amount of particles is labeled with a fluorescent dye while the majority of the particles is invisible to the camera. The key to successful measurements is here that the interpolation method developed by Brockmann et al. \citeyear{brockmann2020utilizing} is successfully applied to compensate the effect of remaining refractive index mismatches in the dense suspension.

\section{Experimental set‑up}
\label{sec:1}
\begin{figure}
\includegraphics[trim={0cm 0cm 0cm 0cm},clip,scale=1]{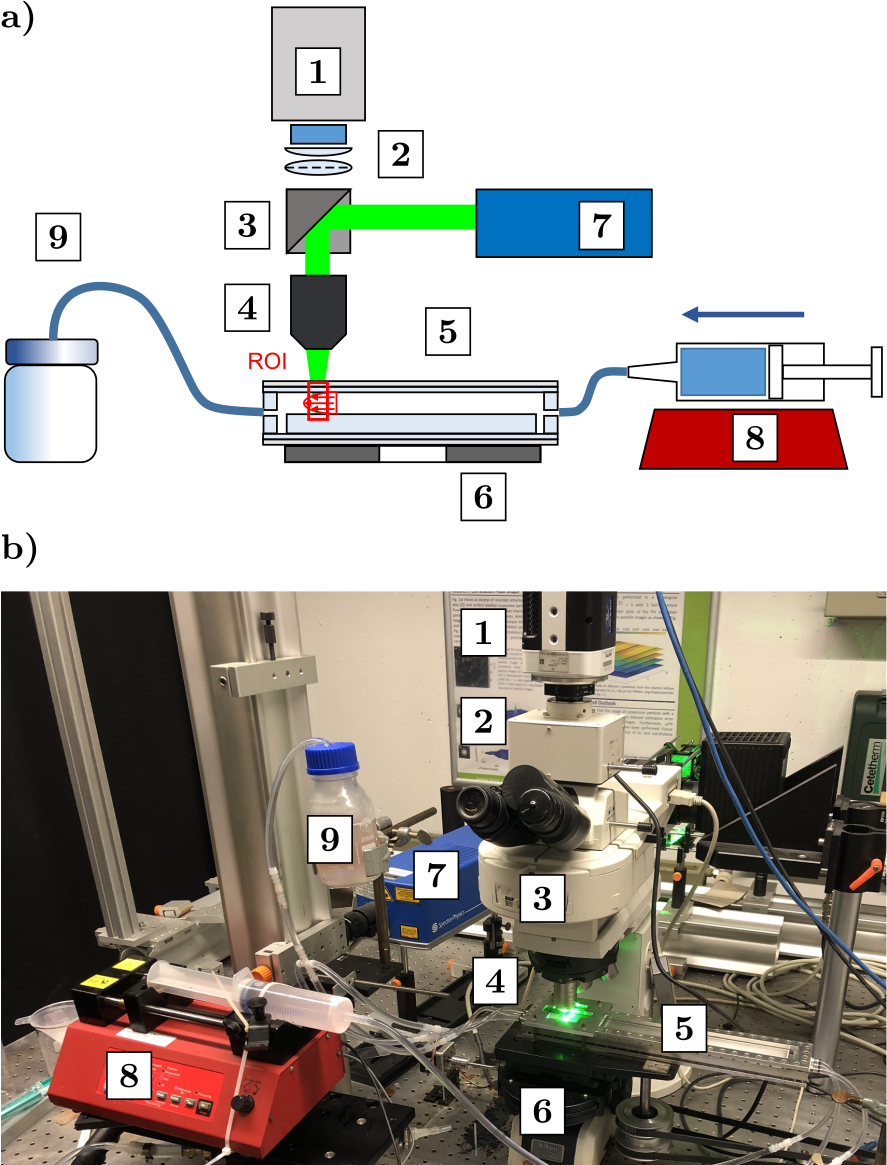}
\caption{Experimental setup: a) Sketch. b) Photograph. 1) Camera, 2) Cylindrical lens and field lens, 3) Dichroic mirror, 4) Microscope objective, 5) Transparent Channel, 6) x,y,z-Traverse, 7) Laser, 8) Syringe Pump, 9) Bottle. }
\label{fig:setup}
\end{figure}

The measurement system consists of a microscope (Nikon Eclipse LV100) illuminated with a \SI{15}{\watt} continuous green laser of \SI{532}{\nano\meter} wavelength. The laser is operated at \SI{1.5}{\watt}. For image-recording a 12-bit, 1280$\times$800 pixel CMOS high-speed camera (Phantom Miro Lab 110, Vision Research) with \SI{20}{\micro\meter} pixel size is used. A shematic drawing of the experimental setup is shown in Fig. \ref{fig:setup}a. 
Measurements are performed with a Nikon Cfi60 objective lens of $M$=10$\times$ magnification. To introduce astigmatism, a cylindrical lens with a focal length of $f_{cyl}$=\SI{200}{\milli\meter} is placed in front of the camera sensor, generating two spatially separated focal planes with a measured distance of approximately \SI{192}{\micro\meter}.
The basic configuration has been already used by Brockmann et al. \citeyear{brockmann2020utilizing} with a bright filed illumination. For the present experiments a plane channel with a cross sectional area of h$\times$w=2.55$\times$\SI{30}{\square\milli\meter} and a length of \SI{300}{\milli\meter} was realized. Velocity profiles were measured \SI{150}{\milli\meter} downstream of the channel entrance. The flow is generated with a high pressure syringe pump (LA-800, Landgraf HLL GmbH) and a \SI{100}{\milli\liter} syringe (Braun GmbH).
For obtaining a density and refractive index matched suspension with $d_p=\SI{60}{\micro\meter}$ PMMA particles we have used the receipe proposed by Bailey and Yoda \citeyear{bailey2003aqueous} with a ternary mixture of 24.85wt\% water, 36.03wt\% glycerin and 39.12 wt\% ammonium thiocyanate that has a refractive index of $n_{\text{RIM}}=1.4867$, a density of $\rho_{\text{RIM}}=\SI{1.19}{\gram\per\cubic\meter}$ and a dynamic viscosity of $\eta_{\text{RIM}}=\SI{4.99}{\centi P}$.

\begin{figure}
\centering
\includegraphics[trim={0cm 0cm 0cm 0cm},clip,scale=1]{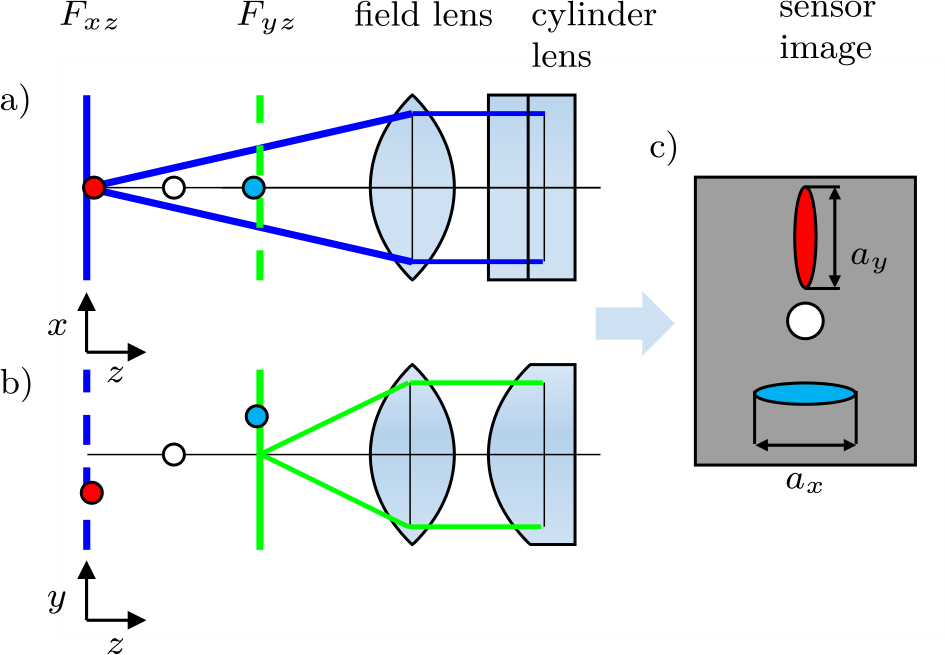}
\caption{Principle sketch of APTV. a) Optical system in $x$-$z$ plane b) Optical system in $y$-$z$ plane c) Resulting image}
\label{fig:principle}
\end{figure}

\begin{figure*}
\includegraphics[trim={0cm 0cm 0cm 0cm},clip,scale=1]{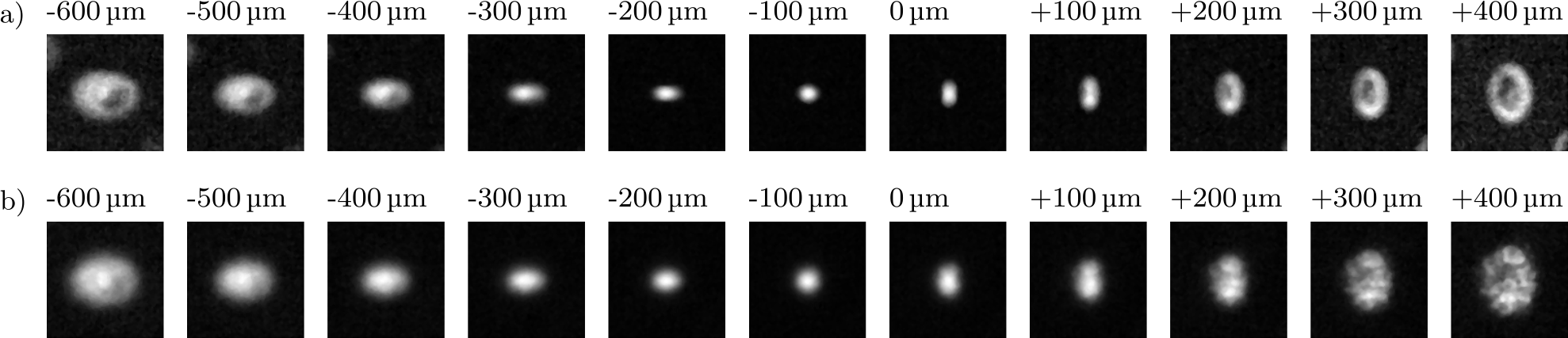}
\caption{ Images of labeled particles for different depth positions $z$ ($d_p$=\SI{60}{\micro\meter}, PMMA, $z$ corrected for refractive index of RIM-liquid ($n_{\text{RIM}}=1.488$)). The particle is located close to $F_{yz}$ at $z\approx-\SI{200}{\micro\meter}$ and located in $F_{xz}$ at $z=\SI{0}{\micro\meter}$. a) Labeled particle in a suspension with $\Phi=0.01\%$. b) Labeled particle in a suspension with $\Phi=19.9\%$.}
    \label{fig: dense_ast_images}
\end{figure*}

\section{Calibration procedure}
\label{sec:calibration_procedure}

\begin{figure}
\centering
\includegraphics[trim={0cm 0cm 0cm 0cm},clip,scale=1]{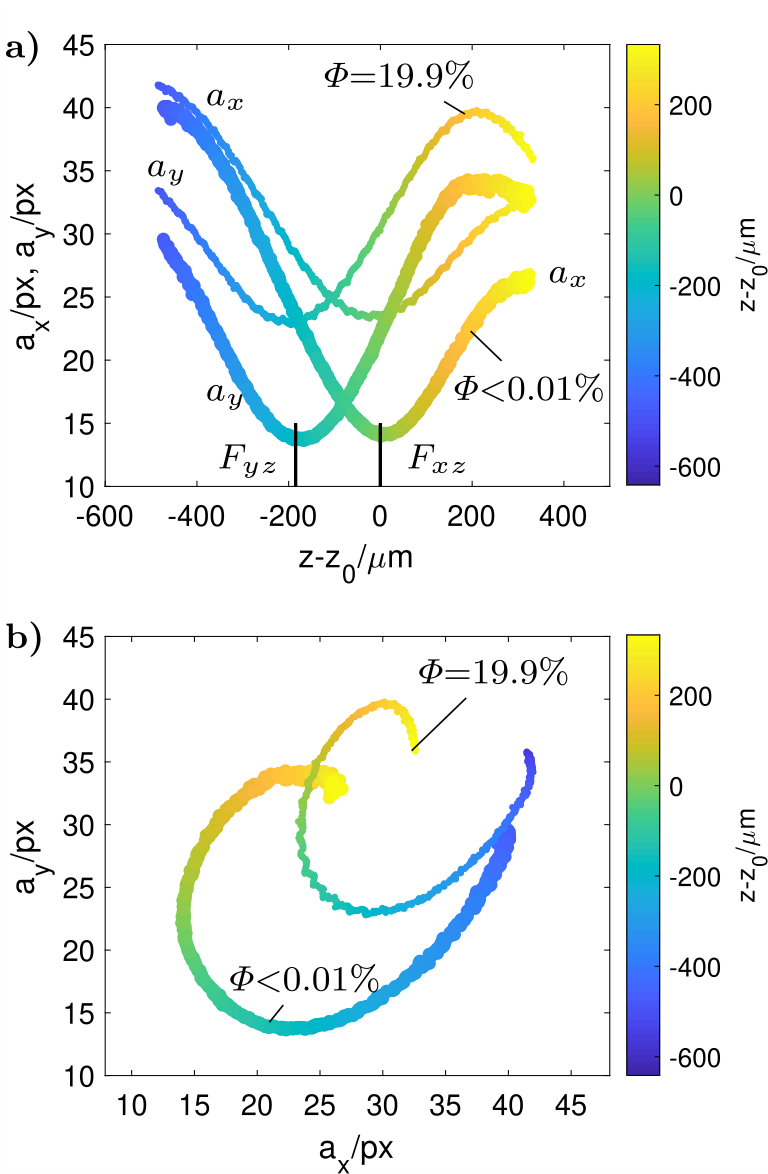}
  \caption{
  Calibration data of $d_p=\SI{60}{\micro\meter}$ particles in RIM-Liquid for $\Phi$<$0.01\%$ (large dots) and $\Phi$=$19.9\%$ (small dots) ($M=10\times$). The position where $a_x$ is minimum ($F_{xz}$) is taken as reference position $z_0$. a) $a_x$ and $a_y$ as function of $z-z_0$ b) $a_y$ as function of $a_x$.
  }
    \label{fig: CAL_E_RIM}
\end{figure}

In APTV, typically a cylindrical lens is implemented in the optical system which alters the light path such that two spatially separated focal planes $F_{xz}$ and $F_{yz}$ are generated. 
If the particle is located at a $z$ position in the middle of the focal planes the particle's image appears circular and skews into an oblate or prolate ellipsoid when the particle's $z$ position is closer to $F_{xz}$ or $F_{xz}$, respectively (see Fig. \ref{fig:principle} and Fig. \ref{fig: dense_ast_images}).
This shape change of the particle image can be quantified by the length of the horizontal and the vertical axis of the autocorrelated particle image denoted as $a_x$ and $a_y$, respectively. To reconstruct the particle's out of plane position based on the values of $a_x$ and $a_y$ a calibration function is required (Cierpka et al. \citeyear{cierpka2010calibration}).
Here, a 2D Euclidean calibration procedure is presented similar to that of Brockmann et al. \citeyear{brockmann2020utilizing}. The main difference is that we utilize fluorescent particles in the present work, therefore the abberated particle image itself is used here for particle out-of-plane position reconstruction. Furthermore, in contrast to Brockmann et al. \citeyear{brockmann2020utilizing} we do not use the 3D Euclidean calibration procedure within this work, as the distribution of the light intensity is not sufficiently homogenous to improve the accuracy through a 3D procedure, here.
The major steps of the calibration procedure are summarized in the following, where we also outline the differences in the calibration curves for a labeled particle in a dilute ($\Phi<0.01\%$) and in a dense refractive index matched suspension ($\Phi=19.9\%$).
To capture the change of $a_x$ and $a_y$ for different out-of plane positions of particles, labeled and wall attached particles in a suspension of $\Phi<0.01\%$ and $\Phi=19.9\%$ are scanned in steps of \SI{1}{\micro\meter} over a distance of \SI{1000}{\micro\meter} such that the deformation of the particle image is entirely captured. Fig. \ref{fig: dense_ast_images} shows particle images of such a scan of a \SI{60}{\micro\meter} PMMA particle labeled with Rhodamin B and located at the bottom channel wall. A magnification of $M=10\times$ for $\Phi<0.01\%$ (Fig. \ref{fig: dense_ast_images}a) and $\Phi=19.9\%$ (Fig. \ref{fig: dense_ast_images}b) is chosen. The particle shown in Fig. \ref{fig: dense_ast_images}a is submerged in a dilute suspension where only labeled tracer particles of $d_p=\SI{60}{\micro\meter}$ are present in the RIM-liquid such that $\Phi<0.01\%$. The change of $a_x$ and $a_y$ is determined as a function of $z-z_0$. Results are displayed in Fig. \ref{fig: CAL_E_RIM}a (large colored dots). The coefficient $c_a$ at which the isolines of the autocorrelation map are extracted to measure $a_x$ and $a_y$ will be denoted as autocorrelation coefficient as introduced and described in Brockmann et al. \citeyear{brockmann2020utilizing}. In Fig. \ref{fig: CAL_E_RIM}b $a_y$ is plotted as a function of $a_x$. This presentation poses the base for the Euclidean calibration method as developed by Cierpka et al. \citeyear{cierpka2010calibration} and adapted in Brockmann et al. \citeyear{brockmann2020utilizing}. 
The labeled particle displayed in Fig. \ref{fig: dense_ast_images}b is submerged in a suspension of RIM-liquid and additional unlabeled particles such that the total volume fraction is $\Phi=19.9\%$. In fact, the suspension was at rest for 24 hours, such that the particle shown in Fig. \ref{fig: dense_ast_images}b is covered with appoximately 11 layers of unlabeled, transparent PMMA particles.  
From Fig. \ref{fig: dense_ast_images}a and b it is obvious that the images of the particle in a suspension of $\Phi=19.9\%$ (Fig. \ref{fig: dense_ast_images}b) are slightly blurred and also exhibit a speckle pattern for relative out-of plane positions ranging between $\SI{200}{\micro\meter}<z-z_0<\SI{400}{\micro\meter}$. The blurriness and the speckle patterns are a result of slight deviations of the refractive index of individual transparent particles and the RIM liquid. The refractive index deviation of transparent particles which are located in the optical path between the labeled calibration particle and the objective, induces these distortions of the particle image affecting the values of $a_x(z-z_0)$ and $a_y(z-z_0)$. These deviations become evident in Fig. \ref{fig: CAL_E_RIM}a (small colored dots) and Fig. \ref{fig: CAL_E_RIM}b (small colored dots), respectively. Obviously the volume fraction affects the $a_x$ and $a_y$ values and particles displayed in Fig. \ref{fig: dense_ast_images}a and b can not be treaded with the same calibration curve.

\begin{figure*}
\centering
\includegraphics[trim={0cm 0cm 0cm 0cm},clip,scale=1]{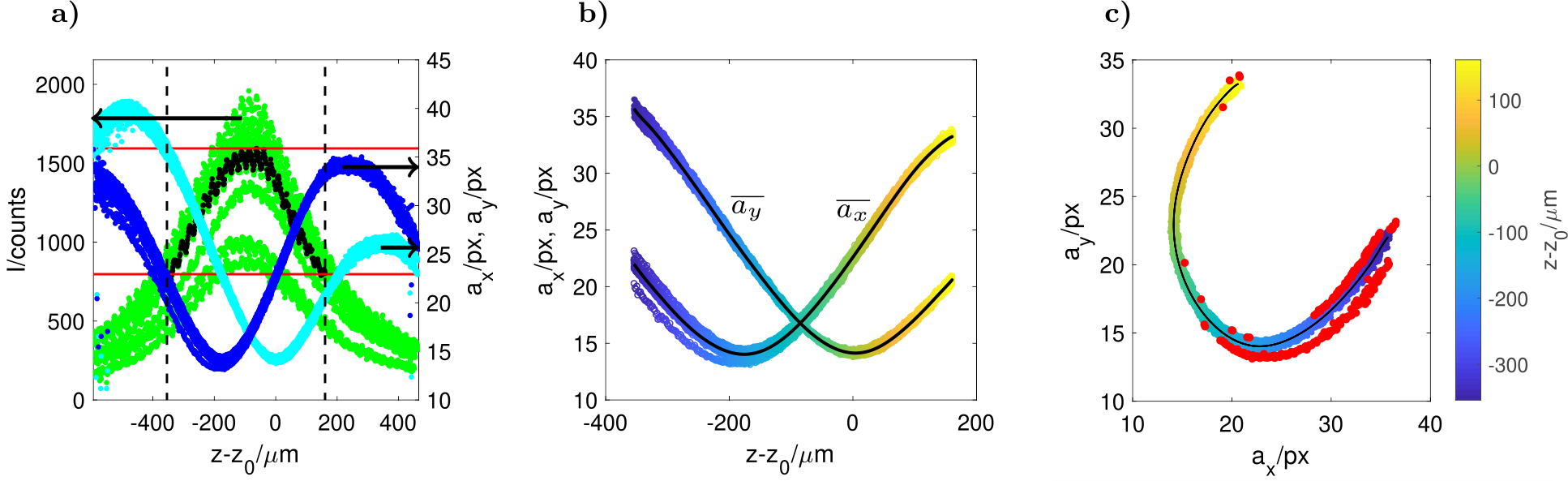}
  \caption{
  Procedure of generating a calibration function  ($c_a=0.5547$, $c_I=0.5$, $c_D=2$, $M=10\times$, $d_p=\SI{60}{\micro\meter}$, $\Phi<0.01\%$). Scale of colormap in b and c is given in c. $z-z_0$ data is corrected for the refractive index of the RIM-liquid ($n_{\text{RIM}}=1.488$). a) Selecting $z-z_0$ range of scattered data by light intensity $I$ (light blue dots=$a_x$, dark blue dots=$a_y$, green dots=$I$, black dots=$\overline{I}$). b) Fitting polynominals of degree 14 to $a_x$ and $a_y$ (black line=polynomials $\overline{a_x}$, $\overline{a_y}$). c) Reconstruction of $z-z_0$ of scattered $a_x$-$a_y$ data (colored dots) by Euclidean distance (black line=polynomials, red dots=outliers).
  }
    \label{fig: gen_cal_func}
\end{figure*}

To obtain a proper calibration function for both the dilute and the dense suspension the previously described calibration scans are repeated for several labeled particles randomly distributed over the field of view. This results in a data set of $a_x$, $a_y$ and $I$ as function of $z-z_0$ for several particles.
An exemplary set of such scattered $a_x$, $a_y$ and $I$ data for $d_p=\SI{60}{\micro\meter}$ particles submerged in a suspension at $\Phi<0.01\%$ is given in Fig. \ref{fig: gen_cal_func}a.
In a first step the median of $I$ as a function of $z-z_0$ is computed, denoted as $\overline{I}$. For the following steps only data points associated with $I \ge c_I \cdot \overline{I}_{\text{max}}$ are considered, where $c_I$ is defined as the intensity coefficient. The value of $c_I \cdot \overline{I}_{\text{max}}$ will be hereafter referred to as intensity threshold $I_{thr}$. For further information on $c_I$ the interested reader is referred to Brockmann et al. \citeyear{brockmann2020utilizing}. 
In the following step, a polynomial of 14th order is fitted to the scattered $a_x$ and $a_y$ data as shown in Fig. \ref{fig: gen_cal_func}. When the polynomial fit of $a_y$ (denoted as $\overline{a_y}$) is plotted as a function of the polynomial fit of $a_x$ (denoted as $\overline{a_x}$) the 2D calibration curve is obtained. The $z$-position of a particle can now be reconstructed by assigning its measured $a_x$, $a_y$ values to a point on the calibration curve that is given by the minimum Euclidean distance and then reading out the associated $z-z_0$ value.
Pairs of $a_x$, $a_y$ where the Euclidean distance exceeds a certain threshold are rejected as an outlier. This threshold is referred to as $a_D$ and defined as the mean Euclidean distance of the $a_x$, $a_y$ data of all calibration particles multiplied by the factor $c_D$ as described in Brockmann et al. \citeyear{brockmann2020utilizing}. For the case depicted in Fig. \ref{fig: gen_cal_func} $c_D$ is set to a value of $c_D=2$, resulting in an out-of-plane reconstruction accuracy of $\sigma_z=\SI{4.67}{\micro\meter}$ for $d_p=\SI{60}{\micro\meter}$ particles. The measurement volume depth for the given example equals $\Delta z=\SI{514.84}{\micro\meter}$ for $c_I=0.5$, such that the relative reconstruction accuracy, hence referred to as relative error, equals $\sigma_z/\Delta z=0.90\%$. With $c_I=0.4$ the measurement volume depth is increased to $\Delta z=\SI{606.28}{\micro\meter}$ whereas $\sigma_z$ increases slightly to \SI{4.92}{\micro\meter} resulting in a relative error of $\sigma_z/\Delta z=0.80\%$. 
%

\section{Validation measurements}
\label{sec: Validation}

To validate the applicability of APTV to dense suspensions, measurements are carried out in a plane channel flow with \SI{2550}{\micro\meter} channel height. The suspension is a ternary liquid mixture proposed by Bailey and Yoda \citeyear{bailey2003aqueous} and PMMA particles of diameter $d_p$=\SI{60}{\micro\meter} (Microbeads). 
The channel is filled with the RIM-liquid and a small amount of labeled particles such that $\Phi<0.01\%$.
Due to a slight density variation of the RIM-liquid and the particles, the particles settle to the channel bottom when rested over night, while a few are stuck to the top channel wall.
Settled particles are used to determine the absolute position of the bottom and the top channel wall prior to the experiments. For this, the whole channel is scanned in steps of \SI{1}{\micro\meter}, to record particles that are located at the top and the bottom wall within the field of view. 
The evolution of $a_x$ is used to detect the particle center and thereby the channel walls by considering the particle radius. In fact the particle center is focused in $F_{xz}$ when $a_x$ assumes a minimum. The origin of the scanning coordinate, is set to zero at the channel bottom. 
Hence, a constant suspension flow rate of \SI{20}{\milli\liter\per\minute} is induced by a high pressure syringe pumpe (LA-800, Landgraf HLL GmbH).
A container is used to collect the liquid driven from the syringe through the channel.

\begin{figure}
\centering
\includegraphics[trim={0cm 0cm 0cm 0cm},clip,scale=1]{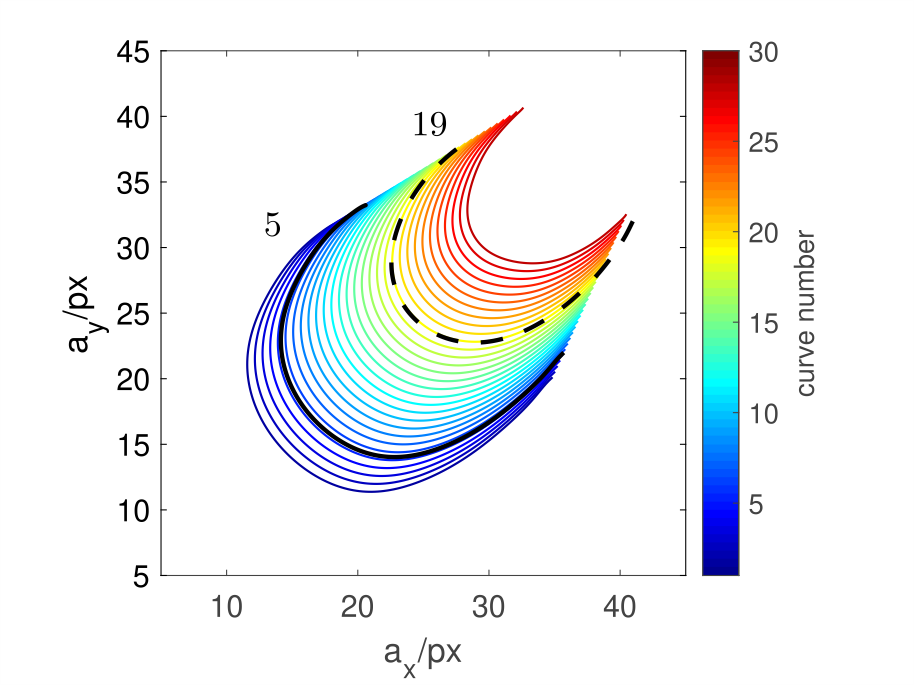}
  \caption{
  Inter- and extrapolated calibration curves as used in Sect. \ref{sec: Validation}. The dotted line equals the calibration curve for $\Phi$=$19.9\%$ (curve number 19), the solid line corresponds to the calibration curve for $\Phi$=$0.01\%$ (curve number 5).  
  }
    \label{fig: interpolation}
\end{figure}

Before the actual flow measurement is started, calibrations measurements are performed with particles located at the bottom channel wall to generate calibration curves for $\Phi<0.01\%$ as shown in Fig. \ref{fig: interpolation} (solid line). Hence, flow measurements for suspensions with six different volume fractions ranging from $\Phi<0.01\%$ to $\Phi=19.9\%$ are performed. 
%
%
After the final measurements at $\Phi=19.9\%$ the setup is rested over night such that particles settle to the channel bottom.
Then, calibration measurements are performed with settled and labeled particles to generate a calibration curve for $\Phi=19.9\%$ as shown in Fig. \ref{fig: interpolation} (dashed line). As already discussed in section \ref{sec:calibration_procedure} both curves differ significantly due to particle image distortions.
%
%
Test calibrations performed on particles located at the top and bottom of the channel revealed that this effect does not occur for a dilute suspension ($\Phi<0.01\%$) in the present case. 
While for low volume fractions no effect can be noticed, for volume fractions larger than $\Phi\ge4.73\%$, we observed that the calibration curve changes with the z-coordinate and hence is an implicit function of the channel height. This is because the number of particles that disturb the light path is higher for labeled particles located closer to the bottom than for labeled particles located closer to the top of the channel. 
The challenge is to find a calibration function which is valid for labeled particles located at any $z$-position in between the bottom and top channel wall and for volume fractions in the range from $\Phi<0.01\%$ to $\Phi=19.9\%$. To solve this problem we adapt the interpolation method developed by Brockmann et al. \citeyear{brockmann2020utilizing}. For this, we interpolate and extrapolate the polynomial coefficients of $\overline{a}_x$, $\overline{a}_y$ based on the calibration curves for $\Phi<0.01\%$ and $\Phi=19.9\%$. In this way, we compute 30 intermediate calibration curves. These are presented as colored lines in Fig. \ref{fig: interpolation}. The extrapolation allows us to generate calibration curves which are even more skewed than the calibration curve for $\Phi=19.9\%$ (see orange to red lines in Fig. \ref{fig: interpolation}). As we will show later these extrapolated curves are required at higher volume fractions.
The $z$-range of all interpolated calibration curves is set to $\Delta z$=\SI{514}{\micro \meter}.
In addition to the calibration curve, which consists of $\overline{a}_y$ as function of $\overline{a}_x$, the threshold for the maximum allowed Euclidean distance $a_D$ is also interpolated linearly.
The out-of-plane position reconstruction uncertainties of particles of $d_p$=\SI{60}{\micro\meter} diameter for $\Phi<0.01\%$ and $\Phi=19.9\%$ are $\sigma_z=\SI{4.67}{\micro\meter}$ and $\sigma_z=\SI{15.64}{\micro\meter}$, respectively.
The channel is scanned in steps of \SI{223}{\micro\meter} during the flow measurements and at each measurement plane 2500 images with a resolution of 512$\times$384 pixel, covering a 1.89$\times$\SI{1.57}{\square\milli\meter} field of view, are recorded at 100fps. In the post processing the image size is reduced to a region of interest of 300$\times$300 pixel. By this, marginal areas of insufficient illumination are reduced and the computation time can be reduced significantly.
After data acquisition, the particle positions and velocities need to be determined. To determine the out-of-plane positions of particles, based on their $a_y$, $a_x$ values, each of the calibration curves displayed in Fig. \ref{fig: interpolation}, is compared with the $a_x$, $a_y$ scatter data from the corresponding measurement planes.
By this we can select an appropiate calibration curve for each measurement plane. To find the best fitting calibration curve, the number of valid $a_y$, $a_x$ pairs that fulfill the Euclidean distance criterion is evaluated as described in Sect. \ref{sec:calibration_procedure}. 
In fact, the curve that yields the largest number of valid particles is considered as a match and selected to determine the $z$-$z_0$ of the $a_y$, $a_x$ pairs in the respective measurement plane. 
\begin{figure*}
\centering
\includegraphics[trim={0cm 0cm 0cm 0cm},clip,scale=1]{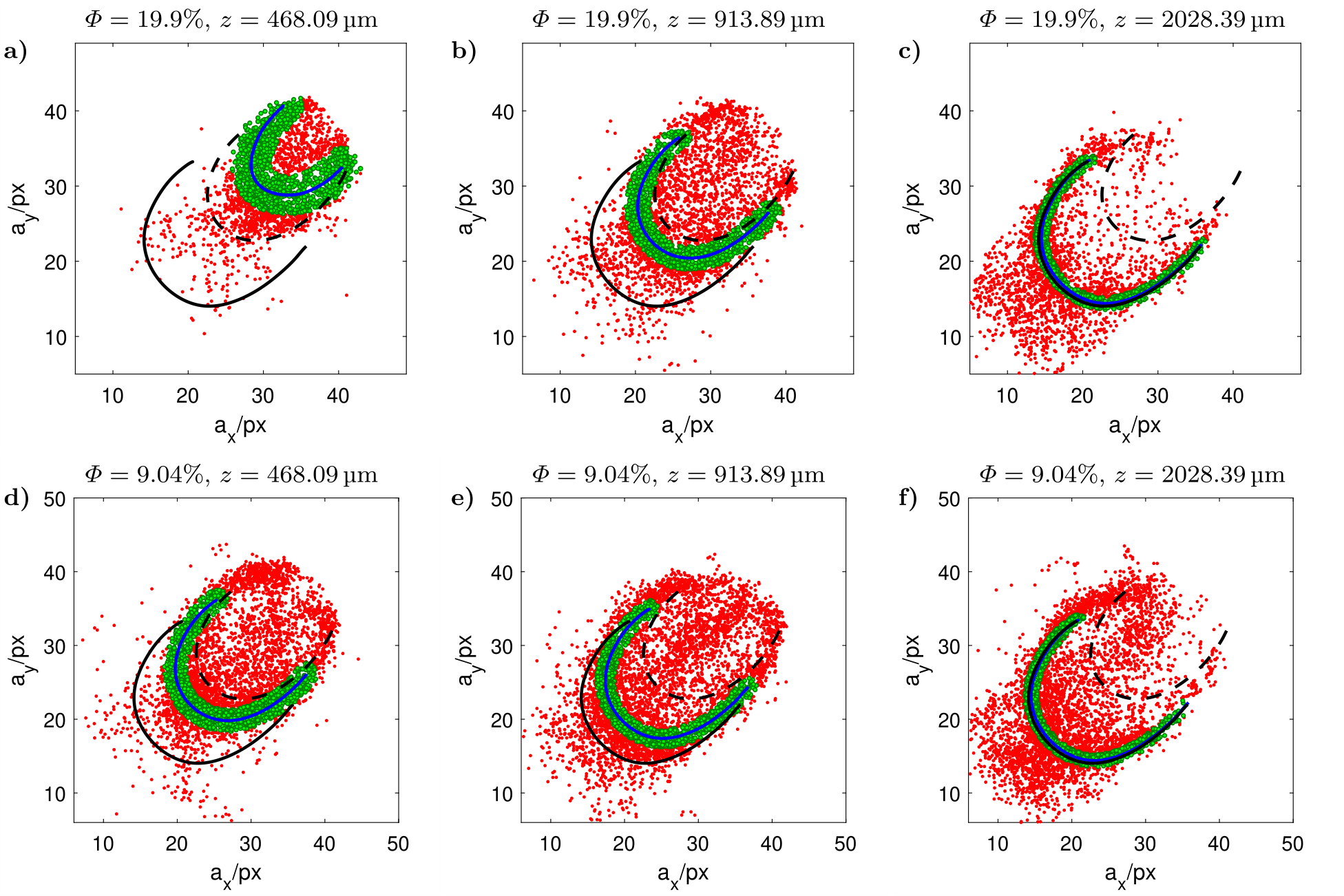}
  \caption{
  Best fitting calibration curve (blue line), valid $a_x$-$a_y$ data (green dots) and rejected $a_x$-$a_y$ data (red dots)  for different measurement planes ($c_a=0.5547$, $c_I=0.5$, $c_D=4$). The solid and the dashed line are the calibration curves as presented in Fig. \ref{fig: interpolation} for a static suspension at $\Phi$=$0.01\%$ and $\Phi$=$19.9\%$, respectively. a) $z=\SI{468.09}{\micro\meter}$, $\Phi=19.9\%$, b) $z=\SI{913.89}{\micro\meter}$, $\Phi$=$19.9\%$, c) $z=\SI{2028.39}{\micro\meter}$, $\Phi$=$19.9\%$, d) $z=\SI{468.09}{\micro\meter}$, $\Phi=9.04\%$, e) $z=\SI{913.89}{\micro\meter}$, $\Phi=9.04\%$, f) $z=\SI{2028.39}{\micro\meter}$, $\Phi=9.04\%$
  }
    \label{fig: intrinsic_check}
\end{figure*}

\begin{figure}
\centering
\includegraphics[trim={0cm 0cm 0cm 0cm},clip,scale=1]{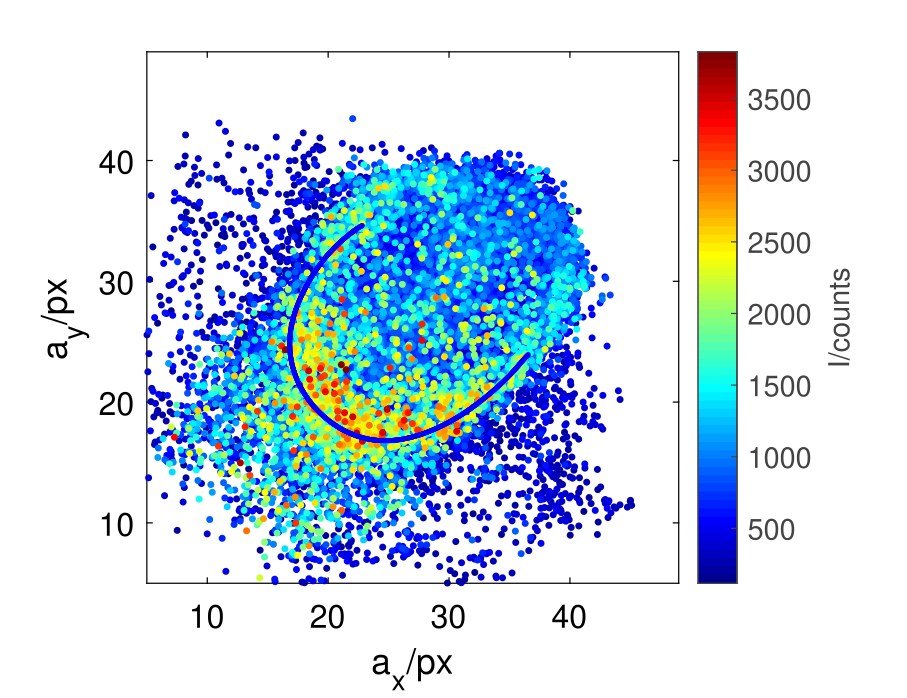}
\caption{
Typical distribution of scattered $a_x$, $a_y$ and $I$ data obtained in one measurement plane ($z=\SI{1136.79}{\micro\meter}$) during the flow measurement for $\Phi$=$12.97\%$. All data gathered in the measurement is displayed. Blue line=best fitting calibration curve.    
}
\label{fig: intensity_curve}
\end{figure}

Figures \ref{fig: intrinsic_check}a-f display the best matching calibration curve (blue solid line) and corresponding valid data points (green dots) and outliers (red dots) as defined in section \ref{sec:calibration_procedure}, for measurement planes located at $z$=\SI{468.09}{\micro\meter}, $z$=\SI{913.89}{\micro\meter} and $z$=\SI{2028.39}{\micro\meter} for $\Phi=9.04\%$ and $\Phi=19.9\%$ particle volume fraction, respectively. 
%
As can be seen in Fig. \ref{fig: intrinsic_check}a it turns out for $\Phi=19.9\%$ at a measurement position of $z=468.09$ an extrapolated calibration curve (blue line) fits best to the measurement data.
The reason for this is, that during the flow measurement the transparent particles are well distributed along the channel height and not settled to a bottom layer as during the calibration. We therefore conclude, that the distortion induced by a static bottom layer of transparent particles on top of a labeled particle is less intense than the distortion created by homogenously distributed transparent particles at the same total volume fraction. Therefore an extrapolation of the calibration curve as displayed in Fig. \ref{fig: interpolation} is needed to capture the calibration curve deformation during the flow measurement.
An interpolation of the maximum Euclidean distance threshold $a_D$ as defined in section \ref{sec:calibration_procedure}) is crucial as the scattering of $a_x$-$a_y$-$I$ data varies along the gap height within the experiments (Brockmann et al. \citeyear{brockmann2020utilizing}). In a suspension flow the scattering increases the closer labeled particles are located to the channel bottom. This can be seen from Figures \ref{fig: intrinsic_check}a-c and d-e where the maximum distance of valid data points (green dots) with respect to the calibration curve (blue line) decreases when the measurement plane is shifted towards the channel top (increasing $z$ values). 
In the present study we use the light intensity of particle images for outlier detection. 
This is essential for the algorithm to reliably pick the best matching calibration curve. The importance of additionaly considering the light intensity as an outlier criterion can be better understood from Fig. \ref{fig: intensity_curve}, where we show the best fitting calibration curve and a typical distribution of $I$ among the $a_x$ and $a_y$ scatter data. As can be seen, the light intensity increases close to the calibration curve. 
For using the light intensity as an additional outlier criterion we inter- and extrapolate the value of the intensity threshold $I_{thr}$ analogous to $a_D$. Data points with $I<I_{thr}$ are rejected. 

\begin{figure}
\centering
\includegraphics[trim={0cm 0cm 0cm 0cm},clip,scale=1]{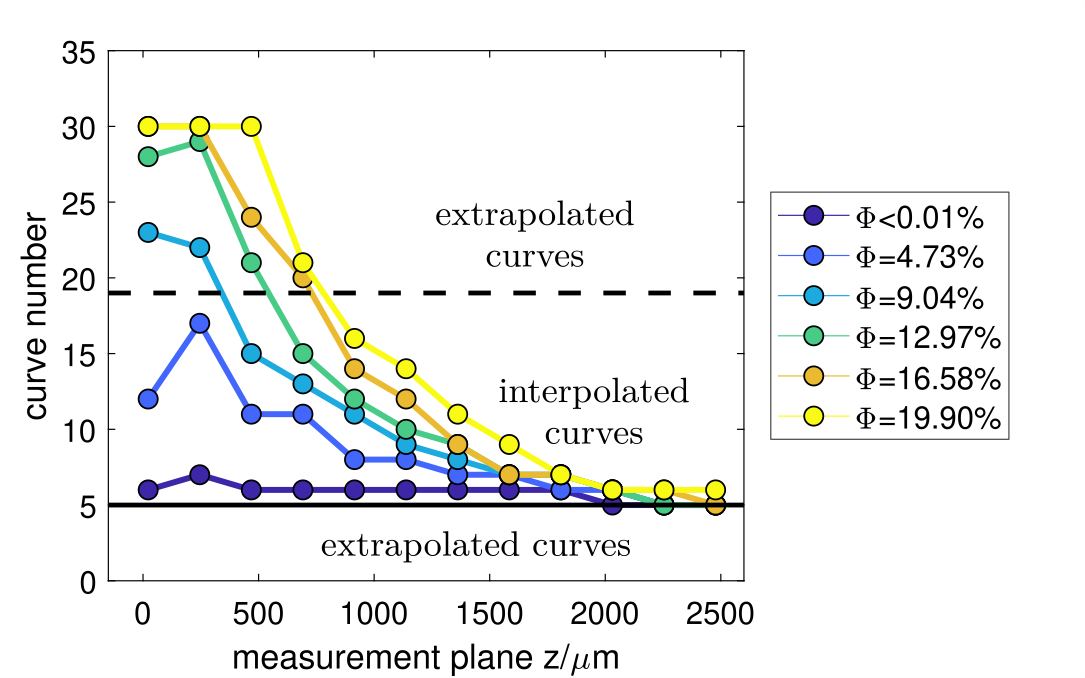}
\caption{Number of best matching calibration curve as function of z-coordinate of measurement plane for the flow measurements presented in Fig. \ref{fig: flow_profiles}. Solid black line=calibration curve for ($\Phi<0.01\%$) (curve number 5 in Fig. \ref{fig: interpolation}), dashed black line=calibration curve for ($\Phi=19.9\%$) (curve number 19 in Fig. \ref{fig: interpolation})
}
\label{fig: best_matches}
\end{figure}

The whole procedure is applied to the $a_x$, $a_y$ and $I$ data of all measurement planes, so that for every measurement plane the out-of-plane positions of the particles are computed. 
Hence, the absolute particle positions can be computed with respect to the channel wall by considering the particle out-of-plane positions and the corresponding measurement plane position.
In Fig. \ref{fig: best_matches} we show which interpolated calibration curve (see Fig. \ref{fig: interpolation}) matches best with respect to the $z$ position of each measurement plane. 
As can be seen for the lowest volume fraction ($\Phi<0.01\%$) the algorithm picks calibration curves 5, 6 and 7 for calculating the out-of-plane positions. For $z>\SI{2000}{\micro\meter}$ and $\Phi<0.01\%$ the best fitting calibration curve matches with the calibration curve for $\Phi<0.01\%$ (number 5). For lower values of $z$, curves 6 and 7 are selected for $\Phi<0.01\%$. However, the differences between the best matching calibration curves for $\Phi<0.01\%$ are small and it is sufficient to consider curve number 5 only for the whole measurement. On the contrary, the curve number changes significantly with increasing $z$ for all flow measurements at higher particle volume fractions ($\Phi\ge4.71\%$). This change becomes more pronounced with increasing volume fraction and appears to be nonlinear as can be clearly seen from Fig. \ref{fig: best_matches}. Obviously for $\Phi\ge9.04\%$ extrapolated calibration curves are required, as the maximum curve number exceeds 19. For $\Phi=19.9\%$ there is a plateau for $z\le\SI{468}{\micro\meter}$, where the curve number is constant. It seems necessary here to extrapolate further to a higher curve number. However, the distortions of the scatter data for $z\le\SI{245}{\micro\meter}$ at the highest volume fraction of $\Phi=19.9\%$ are too strong to capture them by further extrapolation to higher curve numbers. Hence, for measurement planes $z\le\SI{245}{\micro\meter}$  there is a lack of valid data points and the error of the calculated particle position and velocity increases sharply, as will be also discussed hereafter.

\begin{figure*}
\centering
\includegraphics[trim={0cm 0cm 0cm 0cm},clip,scale=1]{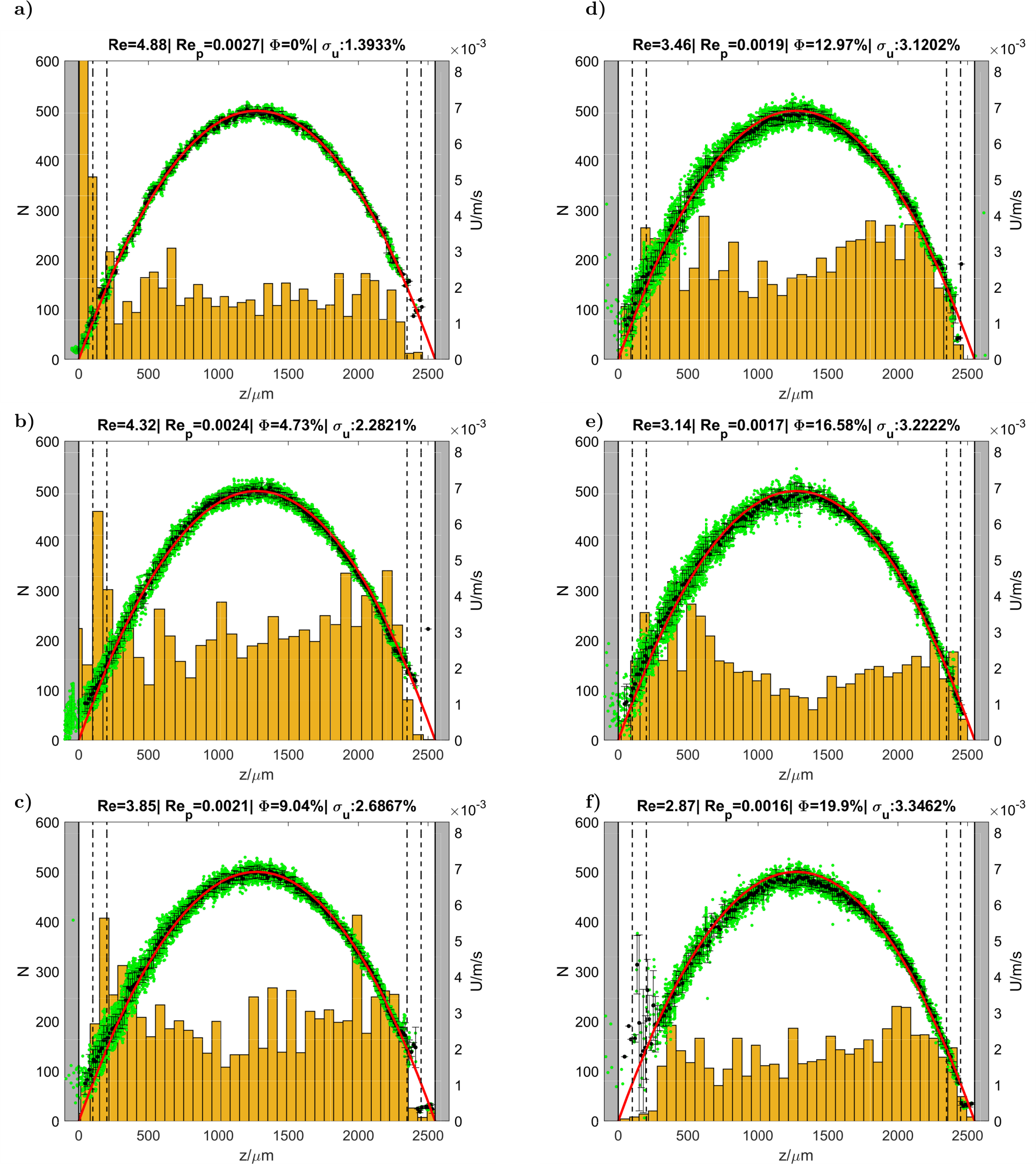}
  \caption{Measured velocity profile and number of valid particles for different particle volume fractions. Brown bar plot = number of valid particles. Red line = analytical velocity profile. Green=scatter data of measured velocity. Black dots with errorbars = averaged velocity. The dashed vertical lines indicate a distance of \SI{100}{\micro\meter} and \SI{200}{\micro\meter} with respect to the wall. a) $\Phi<0.01\%$, b) $\Phi=4.73\%$, c) $\Phi=9.04\%$, d) $\Phi=12.97\%$, e) $\Phi=16.58\%$, f) $\Phi=19.9\%$}
    \label{fig: flow_profiles}
\end{figure*}

Fig. \ref{fig: flow_profiles}a-f show the measured velocity profile and the associated number of valid particles of the plane channel flow for six individual volume fractions between $\Phi\le0.01\%$ and $\Phi=19.9\%$ obtained with the aforementioned procedure. Particle velocities are calculated using a simple nearest neighbors algorithm. For comparison we show the velocity profile for a rectangular channel as given in Shah et al. \citeyear{shah2014laminar} as a red line. The bulk Reynolds number shown in Fig. \ref{fig: flow_profiles}a-f is defined as 
$Re=(U_{\text{Bulk}}D_H \rho_{\text{RIM}})/\nu_{\text{eff}}$ with the effective viscosity estimated as $\nu_{\text{eff}}=\nu_{\text{RIM}}(1+2.5\Phi+5.2\Phi^2)$ according to Batchelor and Green \citeyear{batchelor1972determination}. The particle Reynolds number is estimated as $Re_p=Re(d_p/D)^2$, where $D$ denotes the gap height, as used for example by Shichi et al. \citeyear{shichi2017inertial}.
In general, it can be seen that the measured velocity profile shows a good agreement with the analytical solution for all considered
volume fractions. For higher volume fractions the maximum velocity is slightly lower than the velocity obtained from the analytical solution. This effect could be a result of particle-particle interactions. However the effect is small. 

The uncertainty of determining the in-plane velocity increases from $\sigma_{u}$=1.39\% to $\sigma_{u}$=3.34\% with respect to the maximum streamwise velocity as the volume fraction increases from $\Phi\le0.01\%$ to $\Phi=19.9\%$. The uncertainty for determining the out-of-plane velocity increases from $\sigma_{w}$=9.6\% to $\sigma_{w}$=22.57\% at the same time. 
We assume that the loss of accuracy with increasing volume fraction is related to two different effects. Firstly, the accuracy of determining the particles out-of-plane position decreases with increasing values of $\Phi$ as more layers of transparent particles inducing distortions to the images of the labeled particles. Secondly the accuracy of detecting the in-plane position of the particles centers decreases with increasing volume fractions due to the disturbances induced by the transparent particles. This effect becomes visible in the uncertainty in determining the span-wise velocity component which increases from $\sigma_{v}$=0.50\% to $\sigma_{v}$=1.78\% as the volume fraction increases from $\Phi\le0.01\%$ to $\Phi=19.9\%$. 

The accuracies obtained within this work are lower than those achieved in APTV with small particles and dilute suspensions. This accounts especially for the out-of-plane velocity. For instance Cierpka et al. \citeyear{cierpka2010calibration} obtained uncertainties of $\sigma_{u}$=0.9\% and $\sigma_{w}$=3.72\% of $u_{max}$. We assume that the uncertainty is larger due to slight refractive index mismatches between the RIM-liquid and the transparent particles. 

Regarding the number of detected particles it can be seen from Fig. \ref{fig: flow_profiles}, that there is no obvious lack of data visible among all the planes. Hence, we conclude that the described interpolation procedure is suitable for capturing the shape changes ($a_x$, $a_y$) and the mean Euclidean scattering distance $a_D$ of the calibration curve. In Fig. \ref{fig: flow_profiles}c,d,e  a slight decrease in the number of detected particles towards the middle of the channel becomes evident, which may be related to the measurement procedure and not to a physical phenomena. 
In fact, for all flow measurements the estimated particle Reynolds numbers are small ($Re_p<<0.1$) as can be seen from Fig. \ref{fig: flow_profiles}a-f. Baghat et al. \citeyear{bhagat2009inertial} showed that inertial migration of particles occurs for $Re_p>0.1$. Hence we conclude that such effects can be neglected within our experiments.

Precisely measuring the particle concentration along the channel height is of great importance for the investigation of physical phenomena,  however to developing an APTV based procedure to reliable measure the particle concentration is beyond the scope of this work.


As the particle Reynolds numbers are small and a good agreement could be observed between measured and theoretical velocity profile we conclude that the particle dynamics are negligible here. Thus, particles can approximately be considered as fluidtracers.

\section{Discussion and Conclusion}
\label{sec: Conclusion}

In the present study we show that APTV can be applied to measure the particle dynamics of suspensions of up to 19.9\% volume fraction. Measurements have been performed at six different volume fractions ranging from $\Phi\le0.01\%$ to $\Phi=19.9\%$. 

To make the suspension optically accessible we use a refractive index matched liquid (RIM-liquid) and transparent particles of which just a small portion is labeled with a fluorescent dye. 
Firstly, we study the effect of remaining image aberrations due to slight refractive index mismatches. We observe that slight deviations of the refractive index of individual transparent particles induce optical distortions that result in a shape change of the calibration curve of a labeled calibration particle. Thus, at high volume fractions of transparent particles, images of labeled particles get increasingly distorted the closer they are located to the channel bottom as more transparent particles disturb the optical path.  
To overcome this effect we adapt the interpolation method developed by Brockmann et al. \citeyear{brockmann2020utilizing}. Inter- and extrapolated calibration curves for labeled particles are generated from calibration measurements at $\Phi\le0.01\%$ and $\Phi=19.9\%$ and are related to the measurement data in a best fit procedure. 
In the present study, we used the light intensity as a simple outlier criterion. We found this additional outlier criterion to be crucial for a stable fitting procedure.
Depth reconstruction accuracies of $\sigma_z/\Delta z=0.90\%$ and $\sigma_z/\Delta z=2.53\%$ were achieved for labeled static calibration particles of $d_p=\SI{60}{\micro\meter}$ and a magnification of $M=10\times$ in a suspension of $\Phi\le0.01\%$ and $\Phi=19.9\%$ volume fraction, respectively. 
Ultimately, we validated the interpolation technique successfully by measuring the laminar velocity profile in a rectangular duct with a 2.550$\times$\SI{30}{\square\milli\meter} cross section for six individual volume fractions ranging from $\Phi\le0.01\%$ to $\Phi=19.9\%$. The uncertainty of the measured in-plane velocity was found to be $\sigma_u=1.39\%$ and $\sigma_u=3.34\%$ while the uncertainty for the out-of-plane velocity was $\sigma_w=9.06\%$ and $\sigma_w=22.57\%$ for $\Phi\le0.01\%$ and $\Phi=19.9\%$, respectively. These uncertainties are higher compared to those in APTV with small particles and dilute suspensions.
However, we are convinced that by using a improved illumination technique and the use of a 3D calibration the uncertainties can be further reduced. Furthermore we only used two calibration curves generated at $\Phi\le0.01\%$ and $\Phi=19.9\%$ for our interpolation technique. 
In future works the method could be improved by considering a higher number of volume fractions for generating inter- and extrapolated curves. In fact, preliminary tests (not shown here) indicate that it is benefitial to perform the calibration at solid volume fractions which are actually higher than those used during flow measurements. Finally, to date APTV has been scarcely applied to large fluorescent particles - following studies can help to gain further understanding of how particle size and properties of the optical system affect the measurement accuracy.



\bibliographystyle{apalike}
\bibliography{LITERATURLISTE_exp}

\appendix       

\end{document}